\documentclass{article}
\usepackage{spconf,amsmath,graphicx}
\usepackage{subcaption}
\usepackage{multirow}
\usepackage{array}
\usepackage{arydshln}
\usepackage{booktabs}
\usepackage{color,soul}

\title{Convolutional neural network approach for EEG-based emotion recognition using brain connectivity and its spatial information}
\name{Seong-Eun Moon \quad Soobeom Jang \quad Jong-Seok Lee \thanks{\copyright 2018 IEEE. Personal use of this material is permitted. Permission from IEEE must be obtained for all other uses, in any current or future media, including reprinting/republishing this material for advertising or promotional purposes, creating new collective works, for resale or redistribution to servers or lists, or reuse of any copyrighted component of this work in other works.}}
\address{School of Integrated Technology, Yonsei University \\
	Republic of Korea \\
	\{se.moon, soobeom.jang, jong-seok.lee\}@yonsei.ac.kr} 
\begin{document}
%\ninept
%
\maketitle
\begin{abstract}
Emotion recognition based on electroencephalography (EEG) has received attention as a way to implement human-centric services. However, there is still much room for improvement, particularly in terms of the recognition accuracy. 
In this paper, we propose a novel deep learning approach using convolutional neural networks (CNNs) for EEG-based emotion recognition.
In particular, we employ brain connectivity features that have not been used with deep learning models in previous studies, which can account for synchronous activations of different brain regions. 
In addition, we develop a method to effectively capture asymmetric brain activity patterns that are important for emotion recognition. Experimental results confirm the effectiveness of our approach. 
\end{abstract}

\section{Introduction}
 
Automatic recognition of emotion has been researched actively because it has many useful applications for human-centric services and human-computer interactions. 
Various modalities such as facial expression, affective speech, and gesture have been explored for detecting emotion \cite{Calvo10,Latha16}. In addition, cerebral signals, in particular electroencephalography (EEG), have received much attention in recent years along with the noticeable development of sensing devices, which is expected to contain the comprehensive information of emotion \cite{Alarcao17}. For example, four emotional states (joy, anger, sadness, and pleasure) induced during music listening are classified by using EEG in \cite{Lin10}. 
In \cite{Yazdani12}, it is attempted to conduct binary classification of positive/negative valence, high/low arousal, and like/dislike based on EEG during the watching of music videos. 

While the aforementioned studies use traditional shallow models for the classifiers, deep learning approaches have been introduced recently in the EEG-based emotion recognition. 
Deep belief networks and convolutional neural networks (CNNs) are employed to classify the valence of emotion in \cite{Zheng15} and \cite{Yanagimoto16}, respectively. Unsupervised deep learning methods are also used for automatic feature extraction from EEG signals; for example, autoencoder models are employed in \cite{Liu16} to generate representations of EEG.

In these deep learning approaches, the EEG signal is directly inserted to the models, or some features such as power spectral density (PSD) and differential entropy are first extracted and used as the input of the models. In any case, they ignore the information regarding the spatial arrangement of the EEG electrodes on the scalp.
However, such information provides important patterns in EEG signals, which can be beneficial to compensate for poor signal-to-noise ratios of EEG signals.
In fact, it is shown that %Bashivan et al. \cite{Bashivan16} show that 
two-dimensional image-like representations of PSD features result in improved performance for a mental load classification problem in \cite{Bashivan16}.

Another limitation of the previous studies including the work in \cite{Bashivan16} is that only signals or features from individual electrodes are considered.
In the neuroscience field, however, it has been shown that the relationship between different electrodes is an important clue of brain functions \cite{Friston11,Sakkalis11}.

In this paper, we propose a deep learning approach to incorporate the spatial information of the electrodes and the cross-electrode relationship within CNNs for EEG-based emotion recognition.
We choose to use CNNs because it has the capability to consider the spatial information via the two-dimensional filters in the convolutional layers.
In particular, we focus on using connectivity features extracted from EEG signals as input of the CNNs, which attempts to answer the following issues: ``Which connectivity features are effective for CNN-based emotion recognition using EEG?'' and ``How should the brain connectivity features be represented in order to obtain good recognition performance?''

\section{Proposed method}
\label{sec:recog}

\subsection{Brain connectivity}
\label{subsec:conn}

The brain is a large network of neurons, and synchronous activities of neurons at different regions can provide useful information regarding the neural activity of interest, which is called the brain connectivity. It can be described in three ways \cite{Friston11}. The connectivity between the brain regions that are anatomically connected is called the structural connectivity. The connectivity can be also based on functional integration of separated brain regions, where the directionality can be considered; an undirected dependence is called the functional connectivity, and a causal relationship is called the effective connectivity. 

Therefore, the relationship between brain regions can be described as a brain network where its vertices and edges correspond to brain regions and their connections, respectively. If the edges are weighted, they represent the strength of connections with continuous values. Then, an adjacency matrix can be defined, whose elements are the strength of connections between two electrodes.

\subsection{EEG signal representation}
\label{subsec:feature}

The DEAP \cite{Koelstra12} is employed in this study, which is one of the largest databases and has been popularly used for EEG-based emotion classification. It contains one-minute-long EEG signals measured while subjects were watching affective music videos. Total 32 subjects participated in the experiment, and the EEG signals were recorded for 40 videos by using a 32-channel EEG recording system. Questionnaire results evaluating subjective affective aspects of the videos in terms of arousal, valence, and dominance with a 9-point scoring scale are also provided. 

We consider the problem of binary classification of valence, which indicates whether the emotion is positive or negative. As the subjective questionnaire has a 9-point rating scale, we define the scores ranging from 1 to 5 as the low valence class, and the others as the high valence class. As a result, 55.31\% of the entire data is assigned to the high valence class, and 44.69\% of the data correspond to the low valence class. 

Further processes are applied to the EEG signals for feature extraction. The signals are divided into 3-second-long segments with an overlap of 2.5 seconds, which results in 115 segments of EEG signals per video. The EEG data are split into five clusters randomly, so that the emotion classification is implemented as a five-fold leave-one-cluster-out cross-validation scheme\footnote{Most of the previous studies using the DEAP database implemented subject-wise classification schemes \cite{Gupta16, Daimi14, Zhuang14}. In other words, a part of the data for a subject is used for training and the rest for test, which is repeated for each subject separately. In this case, however, the size of training data is insufficient for training deep CNNs. Therefore, we choose a different classification scheme in this paper. Note that our scheme is more challenging than the subject-wise scheme because EEG patterns varying significantly across different subjects need to be modeled within a single classifier.}. A bandpass filtering is applied to obtain the signals of ten frequency bands: delta (0-3 Hz), theta (4-7 Hz), low alpha (8-9.5 Hz), high alpha (10.5-12 Hz), alpha (8-12 Hz), low beta (13-16 Hz), mid beta (17-20 Hz), high beta (21-29 Hz), beta (13-29 Hz), and gamma (30-50 Hz). Details of extracted features are explained below.

\subsubsection{Activation level of single electrode}
\label{subsubsec:feature-psd}

PSD is commonly employed to describe the activation level of an EEG signal. We compute the PSD of each frequency band from the EEG signal for each electrode by using the Welch's method. 
Total 320 PSD features (32 channels$\times$10 frequency bands) are obtained.

In order to use the PSD values as the input for CNNs, they are transformed to a 32$\times$32 topography based on the location information of the EEG electrodes. For this, the PSD values are allocated to the electrode locations, and the other scalp regions are filled with interpolated values. The topography is obtained for each frequency band, therefore, the EEG signals of a single segment are represented as a 32$\times$32$\times$10 matrix. 

\subsubsection{Connectivity between two electrodes}
\label{subsubsec:feature-conn}

We consider three connectivity features, namely, Pearson correlation coefficient (PCC)%, transfer entropy (TE)
, phase locking value (PLV), and phase lag index (PLI). They are extracted from EEG signals for every pair of electrodes.

PCC describes the linear relationship between two signals $x$ and $y$, and is calculated as:
\begin{equation}
PCC = {cov(x,y) \over \sigma_x \sigma_y},
\end{equation}
where $cov(\cdot)$ indicates the covariance, and $\sigma_x$ and $\sigma_y$ are the standard deviations of the two signals, respectively. PCC ranges from -1 to 1, which correspond to the perfect negative linear relationship and the perfect positive linear relationship, respectively. A PCC value of zero indicates that there is no linear relationship between two signals.

PLV \cite{Lachaux99} represents the phase synchronization between two time series by taking the absolute average of phase differences over temporal windows, which can be written as:
\begin{equation}
PLV = {1 \over N}\left\vert\sum_{n=1}^N e^{j\Delta\phi_n} \right\vert,
\end{equation}
where $N$ is the number of windows, and $\Delta\phi_n$ indicates the phase difference for the $n$-th window. If the two signals are independent, PLV becomes zero. If their phases are perfectly synchronized, PLV becomes one. 	

PLI \cite{Stam07} is also a measure of phase synchronization, but is considered more robust to the common source problem than PLV, which is typically induced by the volume conductance effect or an active reference of the EEG signals. Signs of the phase differences are used for detecting the asymmetry of phase difference distributions, therefore, a peak near zero is ignored. PLI is defined as follows:
\begin{equation}
PLI = \frac{1}{N}\left\vert \sum_{n=1}^N sign(\Delta\phi_n) \right\vert,
\label{eq:pli}
\end{equation}
where $sign(\cdot)$ indicates the sign function.

\begin{figure}[t]
	\centering
	\begin{subfigure}{0.42\columnwidth}
		\includegraphics[width=\textwidth]{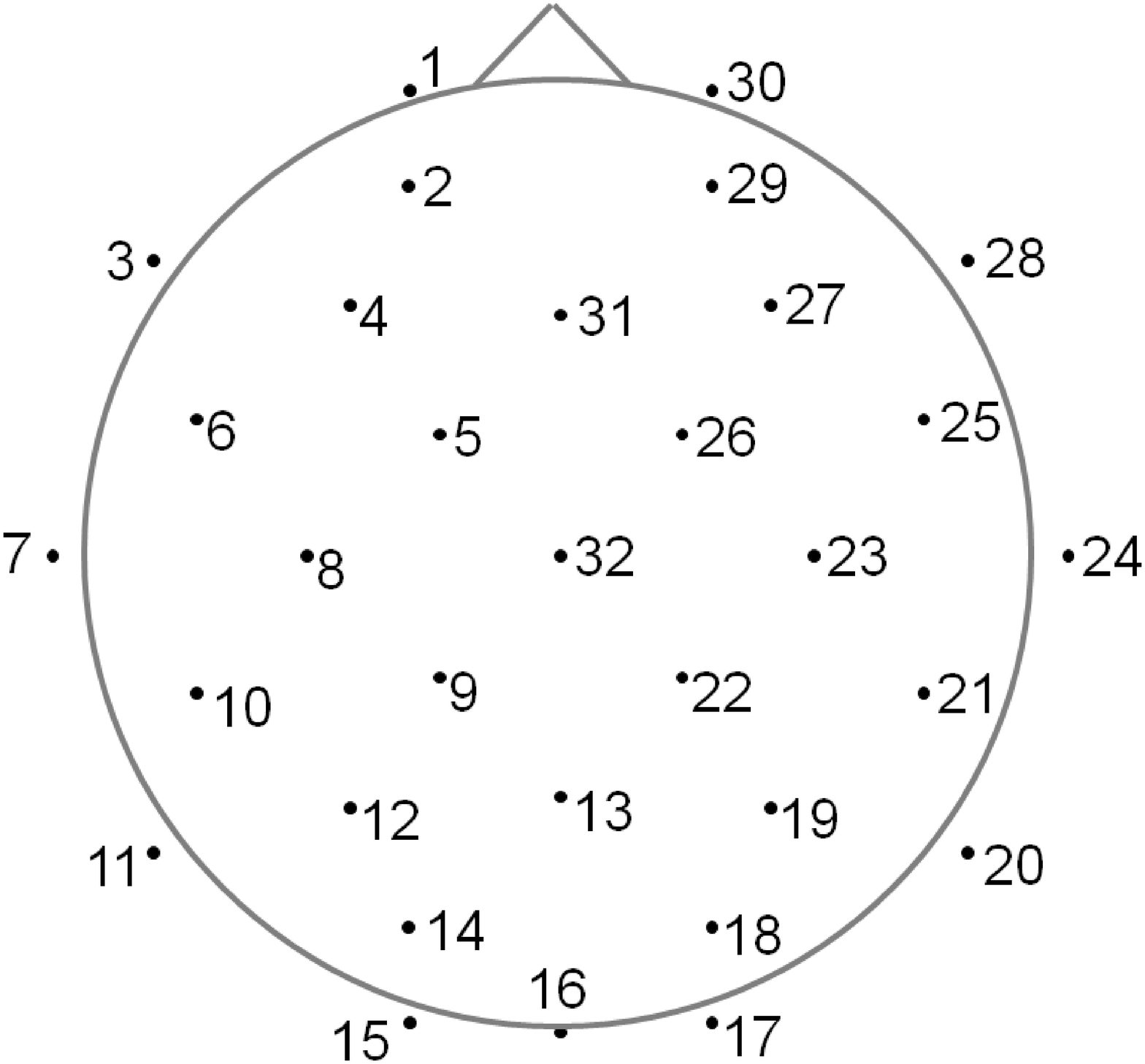}
		\caption{}
	\end{subfigure}
	\begin{subfigure}{0.42\columnwidth}
		\includegraphics[width=\textwidth]{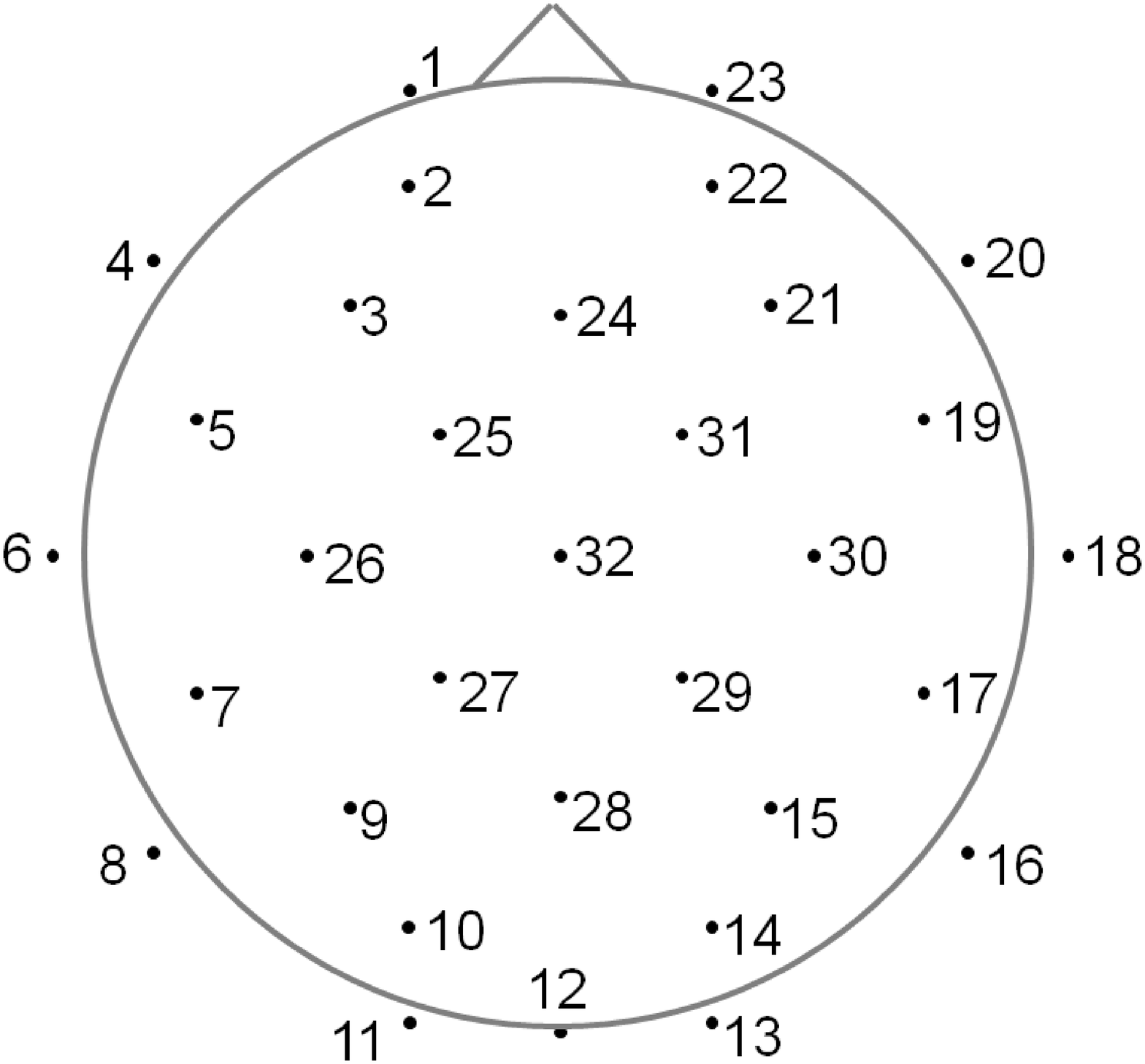}
		\caption{}
	\end{subfigure}

	\vspace{-0.3cm}
	\caption{Ordering methods of the EEG electrodes based on (a) the distance between electrodes with the hemisphere restriction and (b) the distance between electrodes without the hemisphere restriction.}
	\label{fig:ord}
	\vspace{-0.2cm}
\end{figure}

In order to use a set of connectivity features as an input of CNNs, we need to determine how to represent it as a two-dimensional matrix first. 
We transform the connectivity features into a 32$\times$32 connectivity matrix, whose ($i, j$)-th element represents the connectivity feature between the $i$-th and $j$-th electrodes. Here, the order of the electrodes is important because localized filters in a CNN try to learn the patterns of neighboring values in the connectivity matrix. 

We employ two methods that consider the spatial arrangement of the EEG electrodes because the EEG signals of two physically adjacent electrodes tend to be similar due to the volume conductance effect. 
The first ordering method starts from the left frontal electrode (Fp1), goes to the closest electrode in the depth axis of the head within the left hemisphere, moves to the right hemisphere after finishing the left hemisphere, then ends at the center line. The second method also starts from the left frontal electrode but goes to the closest electrode without the hemisphere restriction, which results in a spiral form. Figure \ref{fig:ord} shows the EEG electrodes with numbers assigned depending on the two ordering methods, which will be referred to as the ``dist1'' and ``dist2'' methods in this paper. In addition, we also consider a randomly determined order in order to examine the effect of ordering. 

\subsection{CNN classifiers}
\label{subsec:cnn}

We design three different CNN structures.  
The first CNN has the simplest structure with one convolutional layer and one max-pooling layer (denoted as CNN-2). The second structure (CNN-5) consists of three convolutional layers and two max-pooling layers, i.e., two convolutional layers and one max-pooling layer are added to CNN-2.
The most complex structure (CNN-10) has five convolutional layers and five max-pooling layers, that is, one convolutional layer followed by one max-pooling layer is repeated five times.   
A fully connected layer with 256 hidden nodes is attached in all CNN structures to obtain the output, which is assigned as 0 for the low valence class and 1 for the high valence class. 
All convolutional layers have 3$\times$3 filters, and the number of filters is 32 at the first layer and then becomes twice that of the previous convolutional layer. The rectified linear unit is used as the activation function of the convolutional layers. The max-pooling is implemented for 2$\times$2 patches, and the batch normalization 
is conducted after every max-pooling operation.
We also examined more complex CNN structures, but they did not show any improvement of the classification performance. 

The CNNs are implemented in Theano. They are trained by the Adam algorithm to minimize the loss in terms of the cross-entropy function. The batch size is set to 256. The training is performed using a Tesla K80 GPU.

\section{Results}
\label{sec:result}

\begin{table}
	\centering
	\caption{Classification accuracies.}~\label{tab:resultsingle}
	\vspace{-0.3cm}
	\small
	\begin{tabular}{>{\centering}m{1cm} >{\centering}m{1cm} >{\centering}m{1.5cm} >{\centering}m{1.5cm} >{\centering}m{1.5cm}}% \toprule
		& Ordering Method & CNN-2 & CNN-5 & CNN-10 \tabularnewline \toprule %\midrule
		PSD & & 73.32\% & 80.86\% & 77.90\% \tabularnewline \midrule[0.2pt]
		\multirow{3}{*}{PCC} & random & 93.82\% & 94.44\% & 91.48\% \tabularnewline 
		& dist1 & 93.80\% & 94.17\% & 92.88\% \tabularnewline
		& dist2 & 93.57\% & 94.30\% & 92.68\% \tabularnewline \midrule[0.2pt] %\rule{0pt}{4ex}
		\multirow{3}{*}{PLV} & random & 96.50\% & 97.13\% & 89.93\% \tabularnewline 
		& dist1 & 96.62\% & 97.11\% & 91.30\% \tabularnewline
		& dist2 & 98.58\% & 99.72\% & 90.92\% \tabularnewline \midrule[0.2pt] %\rule{0pt}{4ex}
		\multirow{3}{*}{PLI} & random & 85.00\% & 74.52\% & 59.26\% \tabularnewline 
		& dist1 & 85.03\% & 78.17\% & 61.48\% \tabularnewline
		& dist2 & 84.98\% & 77.48\% & 61.24\% \tabularnewline %\bottomrule %\rule{0pt}{4ex}
	\end{tabular}
	\vspace{-0.2cm}
\end{table}

Table \ref{tab:resultsingle} shows the average accuracy from the five-fold cross-validation for each combination of the feature type, ordering method, and CNN structure. 

When the PSD features are used, the maximum accuracy of 80.86\% is obtained from CNN-5. When we use a SVM classifier having the radial basis function kernel, we obtain an accuracy of 55.42\%. This shows that CNNs can improve the classification performance in comparison to the conventional SVM. However, they show overfitting if their structure is too complex (i.e., CNN-10).

Overall, the connectivity features improve the performance over the PSD features. The best performance (99.72\%) is achieved when CNN-5 is used with PLV matrices formed with the dist2 method. This is a huge improvement in comparison to the performance of the PSD features. Except for a few cases for PLI, the connectivity features always yield better performance than the PSD features. This is consistent with the results of a previous study, which revealed that connectivity features result in better performance than PSD, in particular, when multiple subjects are involved in training and testing phases \cite{Moon15ACMMM}.

\begin{figure}[t]
	\centering
	\begin{subfigure}{3.3cm}
		\includegraphics[width=3cm]{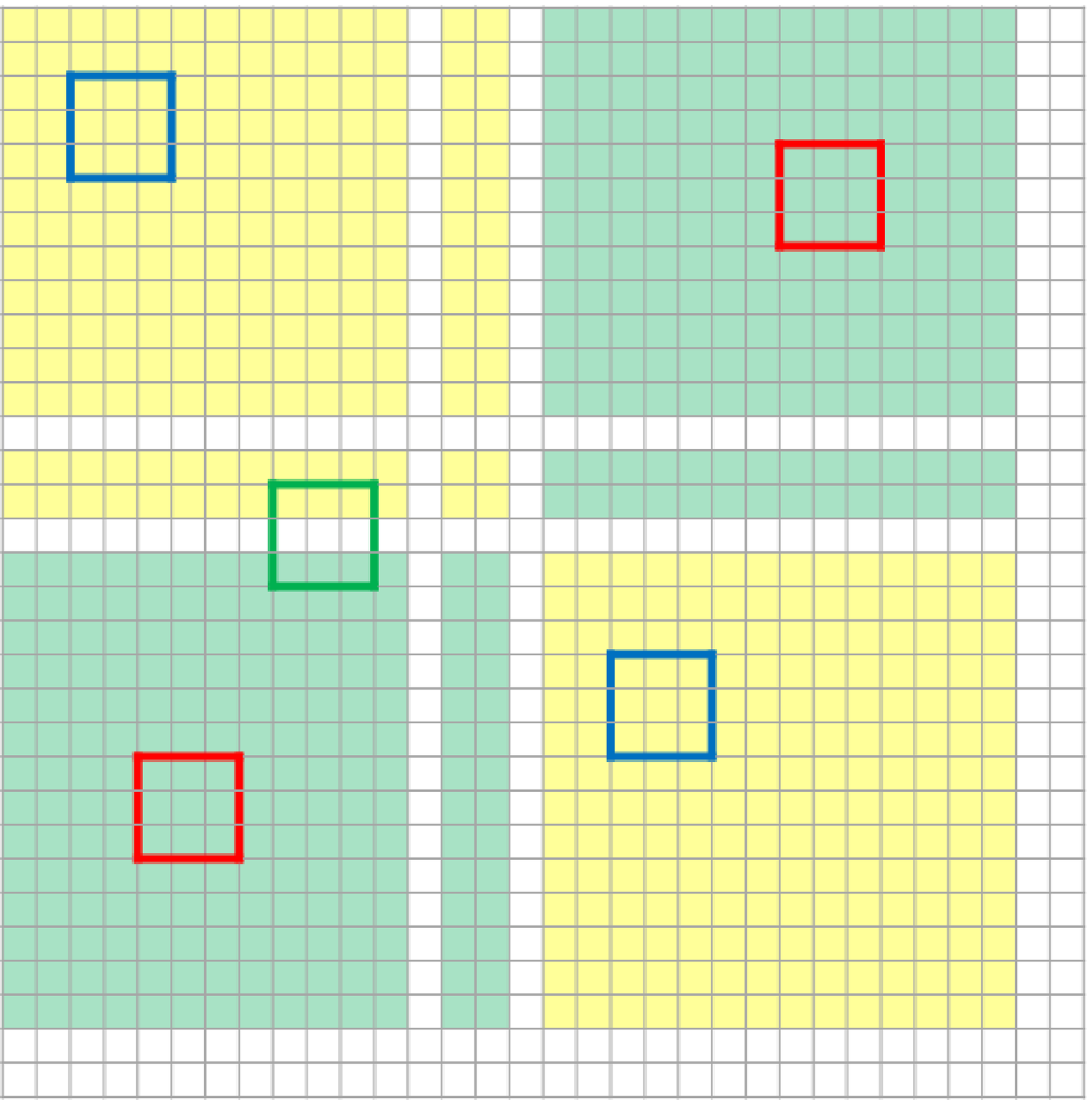}
		\caption{dist1}
		\label{fig:distcomp-dist1}
	\end{subfigure}
	\begin{subfigure}{3.3cm}
		\includegraphics[width=3cm]{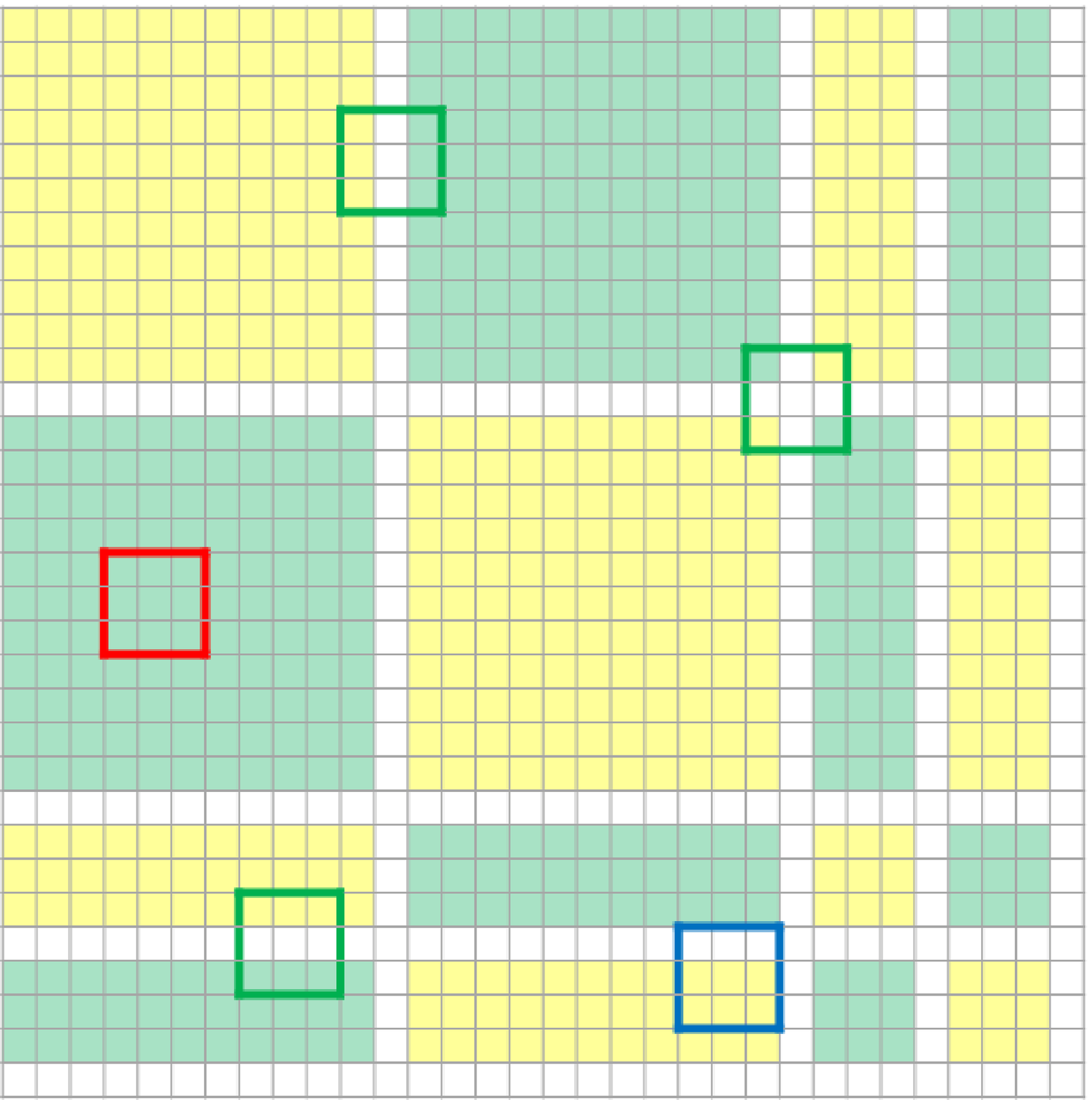}
		\caption{dist2}
		\label{fig:distcomp-dist2}
	\end{subfigure}
	\vspace{-0.3cm}
	\caption{Locations of connectivity features in connectivity matrices. A yellow cell corresponds to a within-hemisphere connectivity, and a green cell indicates a between-hemisphere connectivity. A white colored column or row means that the corresponding electrode is on the center line. Colored 3$\times$3 squares indicate example receptive fields of the filters in the first convolutional layer (see the text). }
	\label{fig:distcomp}
	\vspace{-0.2cm}
\end{figure}

The two ordering methods, dist1 and dist2, show comparable results in most cases, however, the best performance is obtained by dist2. This shows that it is important to place the electrodes from different hemispheres in neighboring columns and rows in the connectivity matrix.

Previous studies consistently reported significant relationship between asymmetry patterns of brain activities and emotional process, based on which the asymmetry has been regarded as an important descriptor of emotion \cite{Coan04, Kim13asy}. 
The asymmetry can be reflected in connectivity features;
roughly speaking, if the brain is activated asymmetrically, the within-hemisphere connectivity becomes relatively high, and the between-hemisphere connectivity becomes relatively low. 

Figure \ref{fig:distcomp} describes the locations of within-hemisphere and between-hemisphere connectivities. 
When the dist1 method is used, most of the receptive fields in the first convolutional layer contain the information of within-hemisphere or between-hemisphere connections (blue or red squares in Figure \ref{fig:distcomp-dist1}, respectively). In this case, between-hemisphere and within-hemisphere connections are rarely included within a single receptive field (the green square in Figure \ref{fig:distcomp-dist1}). In contrast, the combinations of different types of connections are more frequently covered within a single receptive field with the dist2 method (green squares in Figure \ref{fig:distcomp-dist2}). This facilitates effective processing of asymmetric activity patterns by convolutional operations in CNNs. 

According to this argument, it may be thought that the random ordering method should show better performance than the others. However, there is no significant difference between the results of the random and dist2 methods for many cases, and the random method even shows the worst performance for some cases. The physical distance between electrodes is not considered in the random ordering, which means that the connectivity feature values of adjacent elements of a connectivity matrix tend to be quite unrelated. It has the effect that the CNN inputs based on the random ordering method become noisy in comparison to the other ordering methods. In other words, it is hard to have a receptive field covering physically close electrodes in the random ordering method, while such receptive fields are still found in the dist2 method.

When different connectivity features are compared, PLV shows the best performance except for CNN-10, while PCC is effective for the most complex CNN structure. Between the two phase-related features, PLV shows better performance than PLI overall. PLI employs the sign function for robustness as shown in (3), which seems to remove some information useful for classification.

\begin{figure}[t]
	\centering
	\begin{subfigure}{6.5cm}
		\includegraphics[width=\columnwidth]{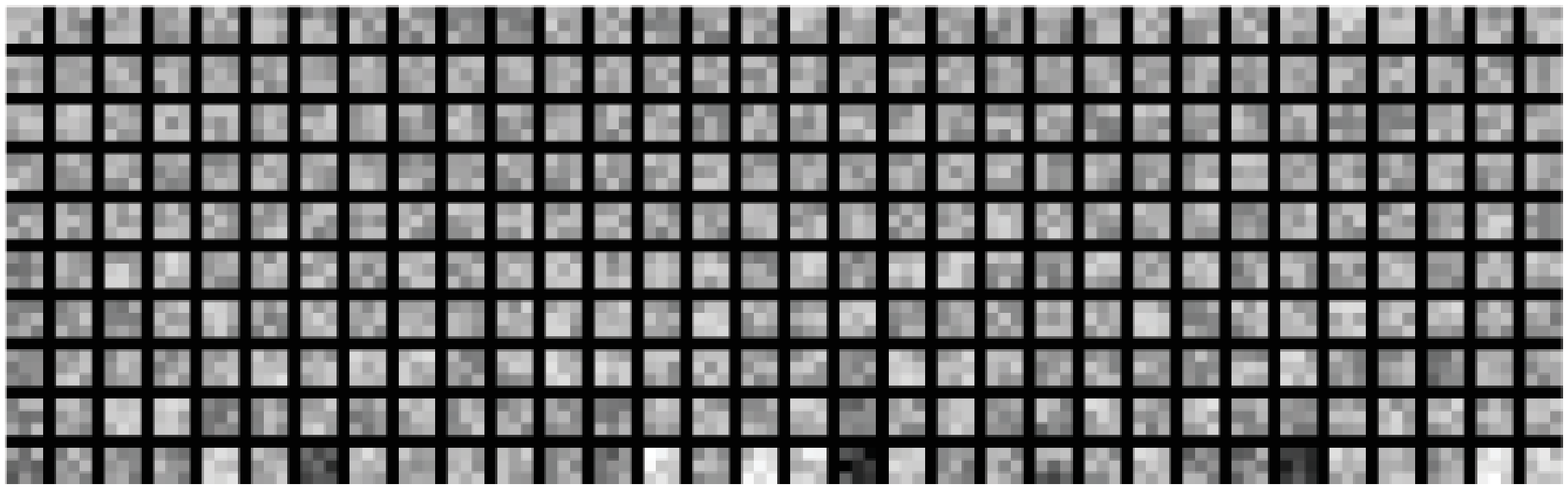}
		\caption{PSD}
		\label{fig:filter-PSD}
	\end{subfigure}
	\begin{subfigure}{6.5cm}
		\includegraphics[width=\columnwidth]{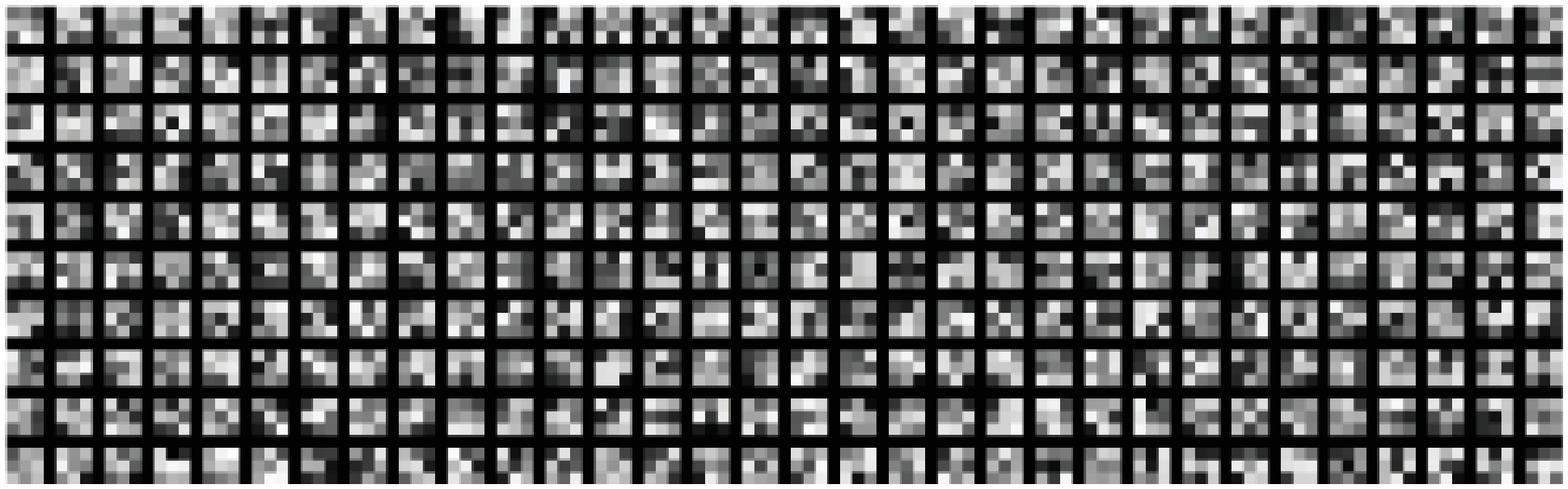}
		\caption{PLV-dist2}
		\label{fig:filter-PLV-dist2}
	\end{subfigure}
	\vspace{-0.3cm}
	\caption{Filter weights of the first convolutional layer of CNN-10.}
	\label{fig:filter}
	\vspace{-0.2cm}
\end{figure}

In order to analyze the feature extraction performance of the proposed model, we examine the filter weights of the first convoluational layer of CNN-5 trained using PLV-dist2 and PSD, which are shown in Figure \ref{fig:filter}. There are 10 input channels and 32 output channels, which correspond to the numbers of the rows and columns of Figure \ref{fig:filter}, respectively. 

Unlike in Figure \ref{fig:filter-PLV-dist2}, filters having almost uniform weight values are observed in Figure \ref{fig:filter-PSD}. These filters do not contribute much in extracting useful features from input data. Conversely, PLV provides rich information to be learned regarding brain activities. 

\section{Conclusion}
\label{sec:conclusion}

We have proposed a new CNN-based approach for emotion recognition using EEG. 
We represented connectivity features as matrices, where we showed the ordering of the electrodes is important for recognition performance.
We also showed that the PLV connectivity features produce the best performance among the three tested connectivity features.

In our future work, we will examine other types of connectivity features. 
In addition, methods to effectively integrate different types of features will be also investigated.

\subsubsection*{Acknowledgments}
This work was supported by Basic Science Research Program through the National Research Foundation of Korea (NRF) funded by the Korea government (MSIT) (NRF-2016R1E1A1A01943283).
%\vfill\pagebreak

\bibliographystyle{IEEEbib}
\bibliography{refs}

\begin{thebibliography}{10}

\bibitem{Calvo10}
Rafael~A. Calvo and Sidney D'Mello,
\newblock ``Affect detection: an interdisciplinary review of models, methods,
  and their applications,''
\newblock {\em IEEE Transactions on Affective Computing}, vol. 1, no. 1, pp.
  18--37, 2010.

\bibitem{Latha16}
Charlyn~Pushpa Latha and Mohana Priya,
\newblock ``A review on deep learning algorithms for speech and facial emotion
  recognition,''
\newblock {\em Journal on Computer Science and Information Technologies}, vol.
  1, no. 3, pp. 88--104, 2016.

\bibitem{Alarcao17}
Soraria~M. Alarcao and Manuel~J. Fonseca,
\newblock ``Emotions recognition using {EEG} signals: a survey,''
\newblock {\em IEEE Transactions on Affective Computing}, vol. PP, no. 99, pp.
  1--20, 2017.

\bibitem{Lin10}
Yuan-Pin Lin, Chi-Hong Wang, Tzyy-Ping Jung, and Tien-Lin Wu,
\newblock ``{EEG}-based emotion recognition in music listening,''
\newblock {\em IEEE Transactions on Biomedical Engineering}, vol. 57, no. 7,
  pp. 1798--1806, 2010.

\bibitem{Yazdani12}
Ashkan Yazdani, Jong-Seok Lee, Jean-Marc Vesin, and Touradj Ebrahimi,
\newblock ``Affect recognition based on physiological changes during the
  watching of music videos,''
\newblock {\em ACM Transactions on Interactive Intelligent Systems}, vol. 2,
  no. 1, pp. 7:1--25, 2012.

\bibitem{Zheng15}
Wei-Long Zheng and Bao-Liang Lu,
\newblock ``Investigating critical frequency bands and channels for {EEG}-based
  emotion recognition with deep neural networks,''
\newblock {\em IEEE Transactions on Autonomous Mental Development}, vol. 7, no.
  3, pp. 162--175, 2015.

\bibitem{Yanagimoto16}
Miku Yanagimoto and Chika Sugimoto,
\newblock ``Convolutional neural networks using supervised pre-training for
  {EEG}-based emotion recognition,''
\newblock in {\em Proceedings of the 8th International Workshop on Biosignal
  Interpretation}, 2016, pp. 72--75.

\bibitem{Liu16}
Wei Liu, Wei-Long Zheng, and Bao-Liang Lu,
\newblock ``Emotion recognition using multimodal deep learning,''
\newblock in {\em Proceedings of the International Conference on Neural
  Information Processing}, 2016, pp. 521--529.

\bibitem{Bashivan16}
Pouya Bashivan, Irina Rish, Mohammed Yeasin, and Noel Codella,
\newblock ``Learning representations from {EEG} with deep
  recurrent-convolutional neural networks,''
\newblock in {\em Proceedings of the 4th International Conference on Learning
  Representation}. 2016, arXiv:1511.06448v3.

\bibitem{Friston11}
Karl~J. Friston,
\newblock ``Functional and effective connectivity: a review,''
\newblock {\em Brain Connectivity}, vol. 1, no. 1, pp. 13--36, 2011.

\bibitem{Sakkalis11}
Vangelis Sakkalis,
\newblock ``Review of advanced techniques for the estimation of brain
  connectivity measured with {EEG}/{MEG},''
\newblock {\em Computers in Biology and Medicine}, vol. 41, no. 12, pp.
  1110--1117, 2011.

\bibitem{Koelstra12}
Sander Koelstra, Christian Muhl, Mohammad Soleymani, Jong-Seok Lee, Ashkan
  Yazdani, Touradj Ebrahimi, Thierry Pun, Anton Nijholt, and Ioannis Patras,
\newblock ``{DEAP}: a database for emotion analysis; using physiological
  signals,''
\newblock {\em IEEE Transactions on Affective Computing}, vol. 3, no. 1, pp.
  18--31, 2012.

\bibitem{Gupta16}
Rishabh Gupta, Khalil Laghari, and Tiago~H. Falk,
\newblock ``Relevance vector classifier decision fusion and {EEG}
  graph-theoretic features for automatic affective state characterization,''
\newblock {\em Neurocomputing}, vol. 174, pp. 875--884, 2016.

\bibitem{Daimi14}
Syed~Naser Daimi and Goutam Saha,
\newblock ``Classification of emotions induced by music videos and correlation
  with participants' rating,''
\newblock {\em Expert Systems with Applications}, vol. 41, no. 13, pp.
  6057--6065, 2014.

\bibitem{Zhuang14}
Xiaodan Zhuang, Viktor Rozgic, and Michael Crystal,
\newblock ``Compact unsupervised {EEG} response representation for emotion
  recognition,''
\newblock in {\em Proceedings of the IEEE International Conference on
  Biomedical and Health Informatics}, 2014, pp. 736--739.

\bibitem{Lachaux99}
Jean-Philippe Lachaux, Eugenio Rodriguez, Jacques Marinerie, and Francisco~J.
  Varela,
\newblock ``Measuring phase synchrony in brain signals,''
\newblock {\em Human Brain Mapping}, vol. 8, no. 4, pp. 194--208, 1999.

\bibitem{Stam07}
Cornelis~J. Stam, Guido Nolte, and Andreas Daffertshofer,
\newblock ``Phase lag index: assessment of functional connectivity from multi
  channel {EEG} and {MEG} with diminished bias from common sources,''
\newblock {\em Human Brain Mapping}, vol. 28, no. 11, pp. 1178--1193, 2007.

\bibitem{Moon15ACMMM}
Seong-Eun Moon and Jong-Seok Lee,
\newblock ``{EEG} connectivity analysis in perception of tone-mapped high
  dynamic range videos,''
\newblock in {\em Proceedings of the 23rd ACM International Conference on
  Multimedia}, 2015, pp. 987--990.

\bibitem{Coan04}
James~A. Coan and John~J.B. Allen,
\newblock ``Frontal {EEG} asymmetry as a moderator and mediator of emotion,''
\newblock {\em Biological Psychology}, vol. 67, no. 1-2, pp. 7--50, 2004.

\bibitem{Kim13asy}
Min-Ki Kim, Miyoung Kim, Eunmi Oh, and Sung-Phil Kim,
\newblock ``A review on the computational methods for emotional state
  estimation from the human {EEG},''
\newblock {\em Computational and Mathematical Methods in Medicine}, vol. 2013,
  no. Article ID 573734, pp. 1--13, 2013.

\end{thebibliography}

\end{document}